\documentclass[reprint,aps,prd,nofootinbib,preprintnumbers,twocolumns,showpacs,10pt, superscriptaddress]{revtex4-1}
\usepackage{epsfig}
\usepackage{graphicx}
\usepackage{amsmath}
\usepackage{amsfonts}
\usepackage{amssymb}
\usepackage{epstopdf}
\usepackage{bm}
\usepackage[usenames,dvipsnames]{color}
\usepackage{hyperref}
\hypersetup{
    colorlinks=true,       
    linkcolor=NavyBlue,          
    citecolor=Magenta}   
\usepackage{soul,xcolor}

\newcommand\ee{\end{equation}}
\newcommand\be{\begin{equation}}
\newcommand\eea{\end{eqnarray}}
\newcommand\bea{\begin{eqnarray}}
\newcommand\mpl{M_{\rm P}}

\newcommand\ie{{\it i.e.}~}
\newcommand\eg{{\it e.g.}~}

\newcommand\eq[1]{Eq.~(\ref{#1})}

\renewcommand\({\left(}
\renewcommand\){\right)}
\renewcommand\[{\left[}
\renewcommand\]{\right]}
\renewcommand{\mpl}{m_{\rm Pl}}

\begin{document}
\preprint{IFIC/14-77}

\setstcolor{red}

\def\thefootnote{\fnsymbol{footnote}}
\title{Phenomenological approaches of inflation and their equivalence}
\author{Lotfi Boubekeur}
\affiliation{ Instituto de F\'isica Corpuscular (IFIC), CSIC-Universitat de Valencia,\\
Apartado de Correos 22085,  E-46071, Spain.}
\affiliation{ Laboratoire de Physique Math\'ematique et Subatomique (LPMS)\\
Universit\'e de Constantine I, Constantine 25000, Algeria.}

\author{Elena Giusarma}
\affiliation{Physics Department and INFN, Universit\`a di Roma ``La Sapienza'', Ple Aldo Moro 2, 00185, Rome, Italy}

\author{Olga Mena} 
\affiliation{
Instituto de F\'isica Corpuscular (IFIC), CSIC-Universitat de Valencia,\\ 
Apartado de Correos 22085,  E-46071, Spain.}

\author{H\'ector Ram\'irez}
\affiliation{ Instituto de F\'isica Corpuscular (IFIC), CSIC-Universitat de Valencia,\\
Apartado de Correos 22085,  E-46071, Spain.}

\begin{abstract}
{In this work, we analyze two possible alternative and model-independent approaches to describe the inflationary period. The first one assumes a general equation of state during inflation due to Mukhanov, while the second one is based on the slow-roll hierarchy suggested by Hoffman and Turner.  We find that, remarkably, the two approaches are equivalent from the observational viewpoint, as they single out the same areas in the parameter space, and agree with the inflationary attractors where successful inflation occurs. Rephrased in terms of the familiar picture of a slowly rolling canonically-normalized scalar field, the resulting inflaton excursions in these two approaches are almost identical. Furthermore, once the galactic dust polarization data from Planck are included in the numerical fits, inflaton excursions can safely take sub-Planckian values. }

\end{abstract}
\pacs{98.70.Vc, 98.80.Cq, 98.80.Bp}

\maketitle

\twocolumngrid
\section{Introduction}

Despite its impressive observational success, the inflationary paradigm  \cite{Guth:1980zm} is still lacking firm confirmation. The crucial missing piece of evidence is the $B$-modes polarization pattern  imprinted in the cosmic microwave background (CMB) at recombination by the inflationary stochastic gravitational waves (GWs). This observable is usually parametrized through the tensor-to-scalar ratio $r\equiv A_t/A_s$, where $A_t$ and $A_s$ are the amplitudes of the primordial tensor and scalar fluctuations\footnote{The scalar and tensor amplitudes are given by 
\bea
\nonumber
A_s(k)&=&A_s \(\frac{k}{k_0}\)^{n_s-1+\frac12\alpha_s\ln\(\frac{k}{k_0}\)}\, \\
\nonumber
A_t(k)&=&A_t \(\frac{k}{k_0}\)^{n_t}\,,
\eea
where $k_0$ is the pivot scale, $n_s$ and $n_t$ are the scalar and tensor spectral indices, respectively, while $\alpha_s\equiv d n_s/d \ln k$ is the running of the scalar tilt.}, respectively, at some pivot scale. The measurement of $r$ is extremely useful because its magnitude directly determines the inflationary energy scale, when the modes observed now were stretched out of the horizon \cite{Lyth:1984yz}. An additional piece of information is given by the scale-dependence of the power spectrum of inflationary GWs. The accurate measurement of this last, would allow to test the so-called standard inflationary  {\it consistency relation} $n_t = -r/8$~\cite{Liddle:1992wi}. However such a measurement might turn out to be very challenging, especially when the amplitude of the $B$-modes is small  \cite{Dodelson:2014exa}. In view of that, the measurement of $n_t$ would entail an additional experimental challenge that might or might not be met in the future generation of CMB observations. One could be led to conclude that perhaps testing the inflationary consistency relation is not the best way to test the inflationary paradigm in its simplest realization \ie  {\it single-field slow-roll inflation}.  An alternative and easier way might be to test the consistency relation in each model of inflation, \ie the relationship between $r$ and $n_s$ in each of the possible scenarios. For instance, the quadratic model $V\propto \phi^2$ predicts $r=-4(n_s-1)$ at first order in slow-roll.  Such consistency relation would be easier to test than the former one  \cite{Creminelli:2014oaa}, given the present and forecasted accuracy in $n_s$ and $r$. However, despite this encouraging feature, this approach is not model-independent, as it assumes explicitly an underlying scenario with a peculiar inflationary potential to  obtain results. On the other hand,  more useful and robust ways to formulate the tests of inflation should ideally be model-independent, capturing the generic features of inflation, without committing to a specific scenario. Said in other words, it would be more appealing to try to work out the inflationary predictions in a model-independent picture where the inflationary potential does not play a crucial role. This will enable us to avoid  treating inflation on a case-by-case basis, but rather in a more general way. In this work, we address this important issue by considering two possible alternative model-independent approaches.   
 
The recent BICEP2 claim of primordial GWs detection  \cite{bicep2,bicep22} underlined the difficulties faced when trying to extract a primordial polarization signal from the ubiquitous galactic foregrounds. 
Despite the general  excitement in the community, soon after these results were released, several studies carried out a re-assessement of the level of galactic dust polarization in the BICEP2 field~\cite{Mortonson:2014bja,Flauger:2014qra}, questioning the cosmological origin of the BICEP2 signal. Recently, the Planck collaboration \cite{Adam:2014bub} has released the results of the polarized galactic dust emission  measurements at 353 GHz in the BICEP2 field. By extrapolating these results to 150 GHz (the frequency where BICEP2 operates) they were able to test the level of dust contamination in the BICEP2 signal.  The Planck analysis suggests that the BICEP2 signal could be, in principle, explained fully in terms a dust component. However, given the large systematic uncertainties on the polarized dust signal, a joint analysis of Planck and BICEP2 data is mandatory, before giving a final interpretation of the BICEP2 signal. 

In a previous study \cite{Barranco:2014ira}, we have shown that using a purely phenomenological parametrization of the inflationary period, the tension between the BICEP2 signal and previous upper bounds on $r$ can be reduced significantly.  In this work, and along the same lines, we explore two alternative approaches to describe the inflationary paradigm, confronting them with the most recent CMB temperature and polarization data. The first approach, considered in Ref.~\cite{Barranco:2014ira}, is the Mukhanov parametrization of inflation \cite{Mukhanov:2013tua}, while the second one is the so-called inflationary Hubble flow formalism \cite{Hoffman:2000ue, Lidsey:1995np}.  We will see that these two approaches appear to be physically equivalent,  because, interestingly, both single out the same regions in the inflationary parameter space. These results suggest that, when analysing inflationary predictions in a model-independent way, one should restrict attention to these regions in the parameter space, as they are the physical ones, ensuring therefore meaningful and robust constraints. 

The rest of the paper is organized as follows. In Sec.~\ref{sec2}, we review the main features  of the Mukhanov parametrization and explain its branches. Next, in Sec.~\ref{sec3}, we introduce the Hubble flow  formalism and analyze its fixed  points.  Section \ref{sec4} is dedicated to the inflaton excursion. In Sec.~\ref{sec5}, we carry out the numerical analyses of both approaches. We end up by drawing our conclusions in Sec.~\ref{sec6}.

\section{Mukhanov parametrization}
\label{sec2}
In Ref.~\cite{Mukhanov:2013tua}, an alternative and model-independent parametrization of the inflationary period was proposed (see Ref.~\cite{Garcia-Bellido:2014wfa, Roest:2013fha} for a similar treatment). Without reference to a specific potential, one can assume the following ansatz 
\be 
{p/\rho} = -1 + {\beta}/{(1+N_e)^\alpha}\,,
\label{ansatz}
\ee
for the equation of state during inflation\footnote{For an extension of the above ansatz, see \eg \cite{Garcia-Bellido:2014gna}.}. In the above ansatz, $\alpha$ and $\beta$ are  phenomenological parameters and are both positive and of ${\cal O}(1)$, and $N_e$ is the number of remaining e-folds to end inflation. In this hydrodynamical picture, the predictions for the scalar tilt and tensor-to-scalar ratio are 
\begin{subequations}
\label{eq:nsr}
\bea
n_s-1&=&-3\frac{\beta}{\(N_\ast+1\)^\alpha}-\frac{\alpha}{N_\ast+1}\,,
\label{eq:ns}\\
r&=&\frac{24\beta}{(N_\ast+1)^\alpha}~,
\label{eq:r}
\eea
\end{subequations}
where $N_\ast$ stands for the number of e-folds at horizon crossing and it usually takes values  around $60$, depending mildly on the reheating details and on $r$ as well. A general prediction of this ansatz is that the tilt is always {\it negative}, regardless of the inflationary scenario, while the tensor-to-scalar ratio can take any value depending on the parameters $\alpha$, $\beta$ and $N_\ast$. Furthermore,  
the running of the tilt $\alpha_s$ is also always {\it negative}. 

The Mukhanov parametrization captures a wide range of models with completely different predictions \cite{Mukhanov:2013tua}. Notice however that this phenomenological description of the inflationary phase is not completely equivalent to the slow-roll picture, as there is no more freedom in the signs of both the tilt and the running.

\subsection{Two Branches}
\begin{figure}[!t]
\hspace*{-0.8cm} 
\includegraphics[width=0.55\textwidth]{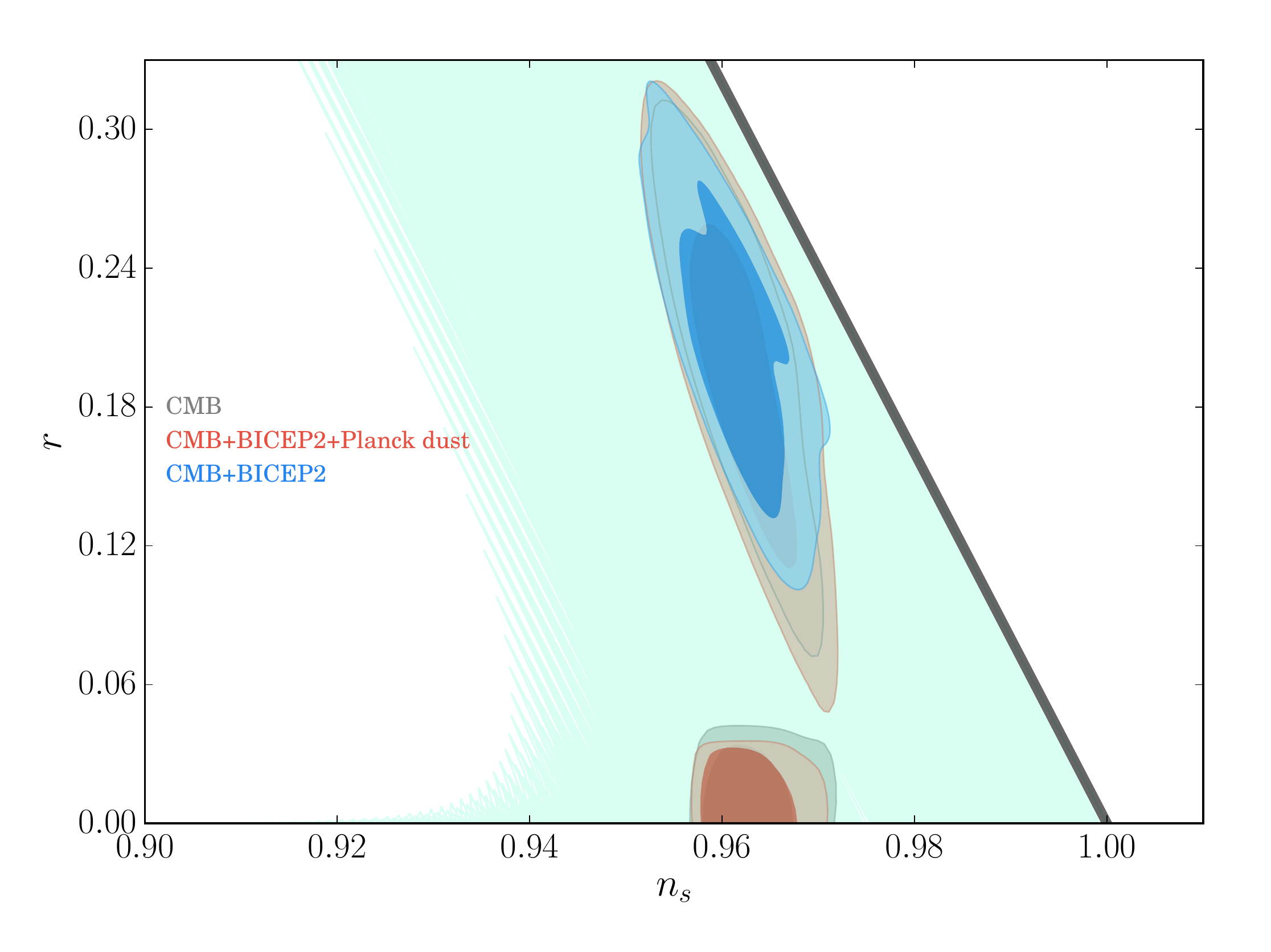}
 \caption{Confidence intervals (68\% and 95\% CL) for the derived parameters $n_s$ and $r$, using the Mukhanov parametrization, from the various data combinations considered in this work. The thick diagonal grey line represents Branch II of  Eqs.~(\ref{eq:brnch}), while the light green area displays the region covered by the Mukhanov parametrization for $N_\ast=40 - 70$.}
\label{fig:fig1}
\end{figure}

As noticed and explained in \cite{Barranco:2014ira}, the Mukhanov parametrization exhibits two distinct branches:

\begin{subequations}
\label{eq:brnch}
\bea
\textrm{Branch I:}~~~&r&\approx 0 ~\textrm{    and ~}n_s\le1\,. \\
\textrm{Branch II:} ~~ &n_s&=1-\frac{r}8\,.
\eea
\end{subequations}
The first branch contains for instance Starobinsky models \cite{Starobinsky:1980te}, while the the second one contains, among other models, the chaotic scenarios\footnote{The natural inflation scenario \cite{Freese:1990rb, Adams:1992bn},  $V(\phi)\propto \[1+\cos(\phi/f)\]$, is captured by the Mukhanov parametrization only for large enough decay constants $f\gtrsim 10 M_P$, which is indeed the regime compatible with observations.} $V(\phi)\propto \phi^n$ \cite{Linde:1983gd}. Because of the presence of these two branches, the observationally preferred value of the scalar spectral index $n_s\simeq 0.96$ will correspond to two different possible values of the tensor-to-scalar ratio, see Fig.~\ref{fig:fig1}. Coming back to the parametrization in terms of $\alpha$ and $\beta$, these two branches are recovered simply as the large and small $\alpha$ limits \ie $\alpha\gg$ 1 and $\alpha \le1$, respectively. Indeed, combining \eq{eq:ns} and \eq{eq:r}, one gets 
\be
\label{nsa}
n_s=1-\frac{r}8-\frac{\alpha}{N_\ast+1}\,.
\ee
From the above expression, and remembering that both $n_s$ and $r$ still depend on $\alpha$, we can easily get the two branches according to whether $\alpha$ is bigger or smaller than $1$.  In principle, the value of the phenomenological parameters $\alpha$ and $\beta$, is unconstrained,  however as discussed in \cite{Barranco:2014ira}, it is sufficient to consider the range $0\le\beta\lesssim 1$ and $0\le\alpha \lesssim 3$. Let us recall some interesting limits of the parametrization \eq{ansatz}. First, the chaotic scenarios $V\propto \phi^n$ correspond to the limiting case $\alpha=1$, regardless of $\beta$. The power $n$ appearing in the potential is given by $\beta=n/6$. Next, the other interesting limiting case is  provided by Starobinski models corresponding to $\alpha=2$ and $\beta=1/2$ in \eq{ansatz}. Finally, the special case $\alpha=0$ corresponds to power-law inflation where the scale factor evolves as $a(t)\propto t^{\frac2{3\beta}}$ and $V\propto e^{-\sqrt{3\beta}\phi/M_P}$. In this scenario, inflation has a graceful exit problem \ie it never ends, and most probably the end of inflation is triggered by an additional field.

\section{The Hubble flow formalism}
\label{sec3}
In this picture, the basic parameter is the Hubble rate $H(\phi)$, and the dynamics can be completely specified without reference to a specific inflaton potential. In this {\it Hamilton-Jacobi} formulation of inflation, starting from $H(\phi)$ and its derivatives, one can construct a hierarchy of slow-roll parameters \cite{Hoffman:2000ue,Lidsey:1995np}. Such parameters start at first order with the usual slow-roll parameters\footnote{As usual, the reduced Planck mass is given by $M_P=(8 \pi G_N)^{-1/2}\simeq 2.43 \times 10^{18}$ GeV. } 
\bea
\epsilon_{\rm H}&\equiv&2 M_P^2\left(\frac{H'(\phi)}{H(\phi)}\right)^2~,\\
\eta_{\rm H}&\equiv& 2 M_P^2\left(\frac{H''(\phi)}{H(\phi)}\right)~.  
\eea
At higher orders, the slow-roll hierarchy is given by 
\bea
^\ell\lambda_{\rm H}\equiv   (2M_P^2)^\ell\, \frac{(H')^{\ell-1}}{H^\ell}\frac{d^{(\ell+1)} H}{d\phi^{(\ell+1)}},\quad \ell\ge2\,.
\eea
These slow-roll parameters obey the infinite system of first order  differential equations 
\bea
{d\epsilon_{\rm H}\over d N}&=& \epsilon_{\rm H}(\sigma_{\rm H}+ 2 \epsilon_{\rm H})\,,\\
{d\sigma_{\rm H}\over d N}&=&-5 \epsilon_{\rm H} \sigma_{\rm H}-12 \epsilon_{\rm H}^2 + 2 \({}^2\lambda_{\rm H} \)\,,\\
\frac{d(^\ell\lambda_{\rm H})}{d N}&=&\[\frac{\ell-1}{2}\sigma_{\rm H} + (
\ell-2)\epsilon_{\rm H}\]{}^\ell\lambda_{\rm H} + {}^{\ell+1}\lambda_{\rm H}\,,
\eea
where the tilt of the scalar spectrum is defined as $\sigma_{\rm H}\equiv 2\eta_{\rm H}-4\epsilon_{\rm H}$. 
Notice that these flow equations are invariant under rescaling the Hubble rate. In principle, they can be integrated to arbitrarily high order in slow-roll \cite{Kinney:2002qn}. In practice,  however, by truncating them at some order $M$; imposing ${}^{M+1}\lambda_{\rm H}$=0,  they become a closed system of differential equations that can be integrated, once a set of initial conditions is specified.

\subsection{Two Fixed Points}
\label{sec:fixp}
By inspection, one can determine the fixed points of the above inflationary flow equations. For instance, truncating at first order, it is straightforward to notice that they exhibit the following fixed points \cite{Hoffman:2000ue}
\bea
\textrm{Fixed point I:}~~~&r&=0 ~\textrm{    and ~}n_s=\textrm{const.} \\
\textrm{Fixed point II:} ~~ &n_s&=1-\frac{r}8\,.
\eea
Fixed point I, can be either stable ($n_s-1>1$) or unstable ($n_s-1<0$) according to the sign of the tilt. We call these fixed points I-a and I-b respectively. The Harrison-Zel'dovich spectrum $n_s=1$  separates these two regions. Remarkably, the fixed points I-b and II of the Hubble flow equations overlap with the two different branches of the Mukhanov parametrization Eqs.~(\ref{eq:brnch}). This is the first main result of this paper. 

\begin{figure}[!t]
\vspace*{-0.6cm}
\hspace*{-0.7cm}
\includegraphics[width=0.57\textwidth,height=8.5cm]{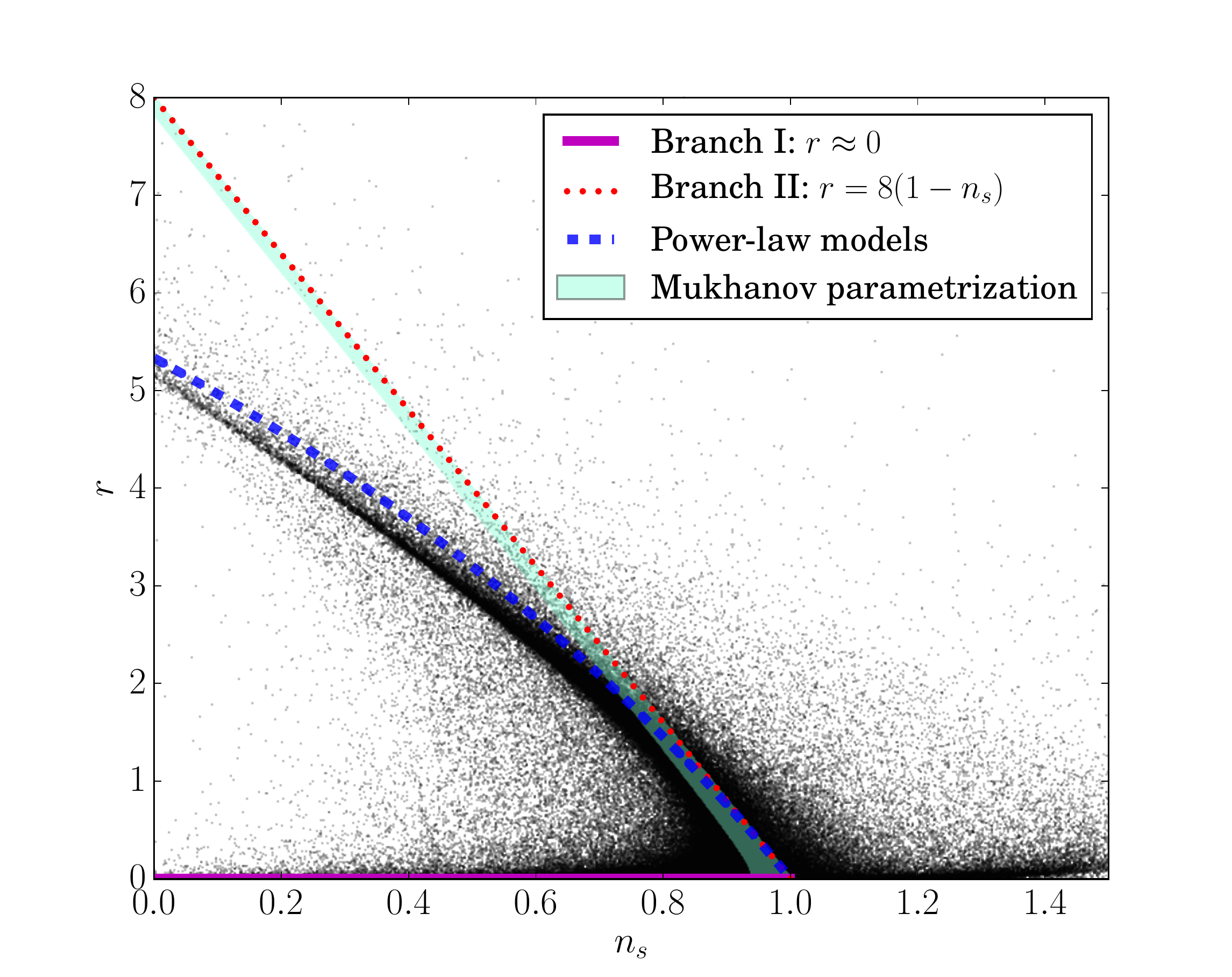}
\vspace*{-0.5cm}
\caption{The results of the Monte Carlo reconstruction using {\tt Flowcode1.0} of $2\times 10^6$ models with wider priors than those shown in Eqs.~(\ref{intervals}). The clustering around Branch I and Branch II is clearly visible. The three different lines refer to the two Branches, together with the attractor for power-law models.}
\label{fig:fig2}
\end{figure}

Considering the full set of equations, the fixed points are given by 
\bea
\textrm{Fixed point I:}~~~&r&=0 ~\textrm{    and ~}n_s=\textrm{const.} 
\label{fix1}\\
\textrm{Fixed point II:} ~~ &n_s&=1-\frac{r}8\times\[\frac1{1-{r}/{16}} \]\,.
\label{fix2}
\eea
The first fixed point, \eq{fix1}, coincides with the first order one, and the stability analysis is the same. However, the second fixed point \eq{fix2} is slightly different and corresponds to power-law scenarios  \cite{Kinney:2002qn},  where $a(t)\propto t^{1/\epsilon_{\rm H}}$. Notice that in this case, $\eta_{\rm H}=\epsilon_{\rm H}$, while $^{\ell+1}\lambda_{\rm H}=\epsilon_{\rm H} (^\ell\lambda_{\rm H})$ for $\ell\ge 2$.  Nevertheless at small $r$, these fixed points coincide; the difference shows only at large $r$, see Figs.~\ref{fig:fig2}.\\

In order to solve the flow equations, we use the publicly  available code {\tt Flowcode1.0} \cite{Kinney:2002qn} that adopts a Monte Carlo approach to reconstruct the inflationary potential. For more details on the methodology, see  \cite{Kinney:2002qn, Easther:2002rw}. For related work using this methodology to obtain cosmological constraints on inflationary models see also \cite{Kinney:2006qm}. We generate a total of $6\times 10^6$ inflationary models by drawing randomly the initial conditions of the slow-roll parameters from the following flat priors~\footnote{For orders $\ell\ge2$, the width of the interval is reduced by a factor of 5 at each order.}
\be
\label{intervals}
\begin{array}{ccl}
N_*&=&\[50,70\]\\
\epsilon_{\rm H}&=&\[0., 0.8\]\\
\sigma_{\rm H}&=&\[-0.1, 0.0\]\\
{}^2\lambda_{\rm H}&=&\[-0.05, 0.05\]\\
{}^3\lambda_{\rm H}&=&\[-0.025, 0.025\]\\
&\cdots&\\
{}^{M+1}\lambda_{\rm H}&=&0.
\end{array}
\ee
As in \cite{Easther:2002rw, Easther:2002rw}, the slow-roll hierarchy is truncated at order $M=8$ and the equations are evolved using {\tt Flowcode1.0}. For illustration, we plot the  results of  reconstructing $2\times 10^6$ inflationary models with wider priors in Fig.~\ref{fig:fig2}.  As noticed in \cite{Hoffman:2000ue}, models cluster around the attractors given by the fixed points. Figure \ref{fig:fig2} clearly shows this feature: in the  $(r, n_s)$ plane, the models populate the regions I-b and II, while the areas  outside these regions are underpopulated. 


\section{The inflaton excursion}
\label{sec4}
\begin{figure}[!t]
\vspace{-0.3cm}
\hspace*{-0.5cm}
\includegraphics[width=0.52\textwidth, height=7.6cm]{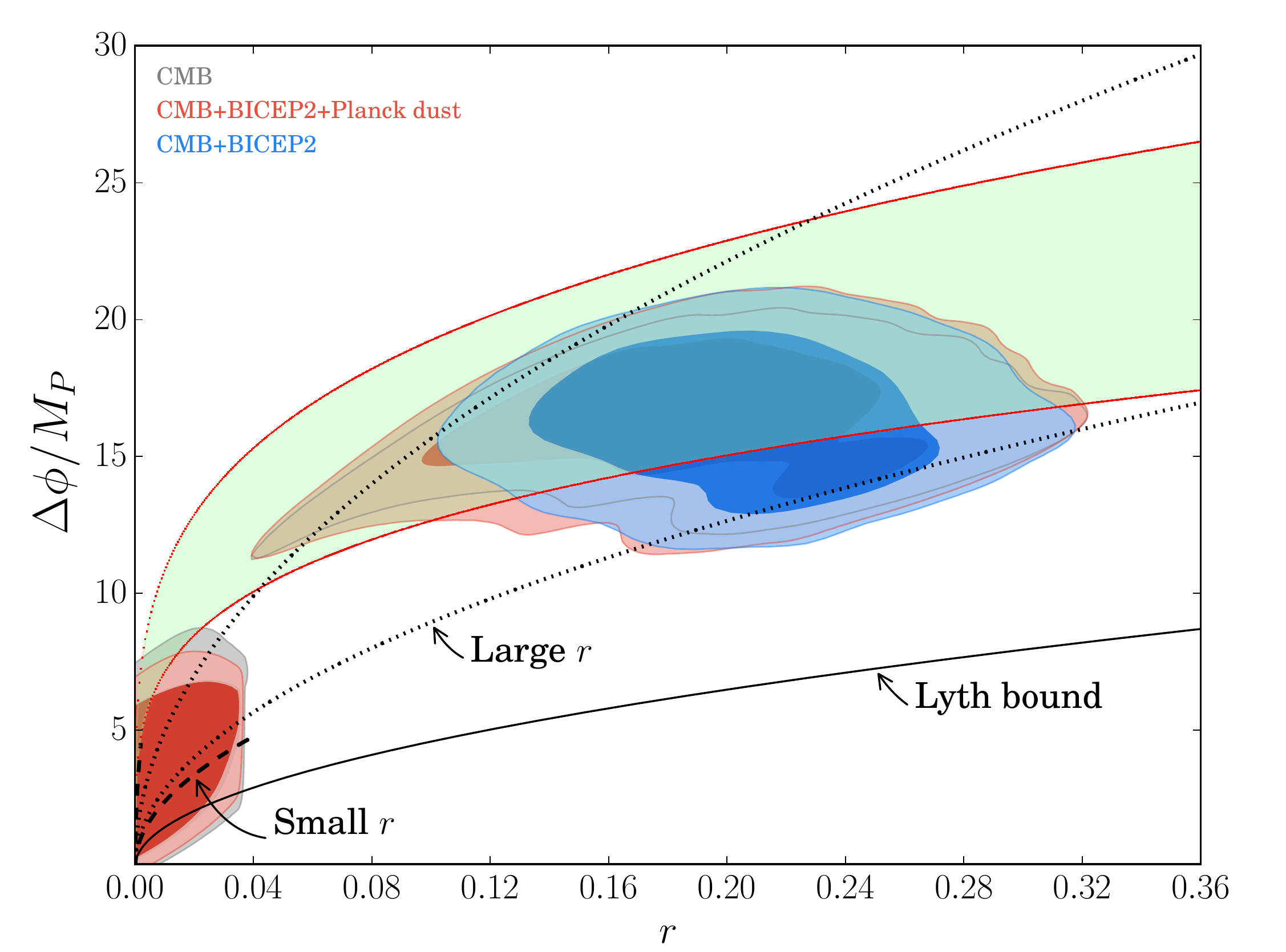}
\vspace{-0.7cm}
\caption{Confidence intervals (68\% and 95\% CL) on the inflaton excursion versus the tensor-to-scalar ratio $r$ from the various data combinations in the Mukhanov parametrization. The light green area represents the theoretical prediction \eq{EM}. The area between the dotted lines refers to the large $r$ limit \eq{eq:fieldlarger}, while the one between the dashed lines refers to the small $r$ limit, \eq{eq:fieldsmallr}.   Finally, the black line stands for the original Lyth bound.  All the regions are computed  for $N_\ast=40 - 70$.}
\label{fig:fig3}
\end{figure}

The Mukhanov parametrization is formulated independently of any scalar field, however one can always recast the dynamics in the inflaton picture \cite{Mukhanov:2013tua}, where inflation is driven by a canonically-normalized scalar field. In slow-roll $\rho\simeq V$, the distance traveled by the inflaton during inflation, \ie the inflaton excursion, can be written in terms of the Mukhanov phenomenological  parameters as
\be
{\Delta \phi\over M_P} = \int_0^{N_\ast} dN \sqrt{3 \beta\over (1+N)^\alpha}\,.
\label{dp}
\ee
For a related recent appraisal of the inflaton excursion see \eg \cite{Garcia-Bellido:2014wfa}. The expression \eq{dp} can be straightforwardly integrated, giving 
\be
\frac{\Delta\phi}{M_P}= \left\{
\begin{array}{l l}
\sqrt{3\beta}\ln(N_\ast+1) &\quad \textrm{ for } \alpha=2\,.\\
\frac{\sqrt{3\beta}}{1-\frac{\alpha}{2}} \[(N_\ast+1)^\frac{-\alpha+2}{2}-1\] &\quad \textrm{ for } \alpha\ne2\,.\\
\end{array}
\right.
\label{eq:field}
\ee
For $\alpha\ne2$, it is useful to consider the small $r$ limit of \eq{nsa}. Recall that CMB data prefers $\alpha> 2$  \cite{Barranco:2014ira}. When $r\ll8/N_\ast$, we can expand around $\alpha=2$, and get 
\be
{\Delta\phi\over M_P}\simeq\sqrt{r\over 8}(1+N_\ast)\[\ln(1+N_\ast) + \frac{(\alpha-2)}{4}\ln(1+N_\ast)^2\]~.
\label{eq:fieldsmallr}
\ee
Figure \ref{fig:fig3} shows the inflation excursion in this limit, for  the range $40<N_\star<70$ and $\alpha=2.6$. Notice that the field excursion in this limit is small, as expected, due to the smaller $r$ in this case. While for the opposite limit, \ie for large $r$ such that $r\gtrsim 8/N_\ast$, $\alpha\simeq 1$ (see \cite{Barranco:2014ira}) and one gets
\be
{\Delta\phi\over M_P}\simeq2 \sqrt{r\over 8}N_\ast\,,
\label{eq:fieldlarger}
\ee
well above the original Lyth bound \cite{Lyth:1996im} (see also \cite{Boubekeur:2005zm, Boubekeur:2012xn}) and in agreement with the predictions for chaotic inflationary scenarios $V(\phi) \propto \phi^n$.  The predictions for the field excursion as a function of $r$ for this regime are also shown in Fig.~\ref{fig:fig3}. Note that, in this case, large field excursions are correlated with large tensor-to-scalar ratios, as expected from the Lyth bound. 

%

In Fig.~\ref{fig:fig3},  we show the {\it derived} inflaton excursion $\Delta\phi/M_P$ versus $r$ in the Mukhanov parametrization arising from our numerical fits to cosmological data, as we shall explain in the next section.  The models cluster around the empirical Efstathiou-Mack relationship\footnote{Notice that here we are using the Planck mass $\mpl=\sqrt{8\pi} M_P\simeq 1.22\times10^{19}$ GeV, instead of $M_P$, in order to compare with the original literature.}~\cite{Efstathiou:2005tq}(see also Ref.~\cite{Verde:2005ff}) 
\be 
\label{efm}
\frac{\Delta\phi}{\mpl}\approx 6 r^{1/4}\,.
\ee  
Such expression has been understood analytically \cite{Boubekeur:2012xn} as the prediction of the {\it quartic hilltop} inflation scenario where $V(\phi)= V_0-\lambda\phi^4/4$. The general prediction for this scenario reads
\be
\frac{\Delta\phi}{\mpl}={N_\ast^{3/4}\over 2 \sqrt{\pi} } \, r^{1/4}\,. 
\label{EM}
\ee

For $N_\ast=60$, \eq{EM} simply reduces to the Efstathiou-Mack relationship, \eq{efm}. Furthermore, \eq{EM} is a special case of the more general hilltop potentials parametrized as $V(\phi)=V_0\[ 1-\lambda_p (\phi/\mu)^p\]$, where $p>2$ and $M_P>\mu>0$. It is straighforward to check that in the Mukhanov parametrization, this corresponds to setting $\alpha=4$. The light green areas in Figs.~\ref{fig:fig3} and \ref{fig:fig4} stand for the prediction given by \eq{EM}, for  $N_\ast$ between 40 and 70.   

\begin{table*}[!t]
\onecolumngrid
\begin{center}
\renewcommand{\arraystretch}{1.4}
\begin{tabular}{c|l|c}
\hline\hline
 Parameter & Physical Meaning & Prior\\
\hline
\hline
$\Omega_{b}h^2$ & Present baryon density &$0.005 \to 0.1$\\
$\Omega_{c}h^2$ & Present Cold dark matter density &$0.001 \to 0.99$\\
$\Theta_s$ & Ratio between the sound horizon and the angular
diameter distance at decoupling&$0.5 \to 10$\\
$\tau$ & Reionization optical depth &$0.01 \to 0.8$\\
$\log{(10^{10} A_{s})}$ & Amplitude of the primordial scalar spectrum &$2.7 \to 4$\\
$\alpha$ & Phenomenological parameter of the Mukhanov parametrization \eq{ansatz}&$0 \to 2.5$\\
 $\beta$& Phenomenological parameter of the Mukhanov parametrization \eq{ansatz}&$0 \to 1$\\
 $N_\star$& Number of e-folds at horizon crossing& $50 \to 70$\\
\hline\hline
\end{tabular}
\caption{Uniform priors on the cosmological parameters used in the {\tt CosmoMC} analyses of the Mukhanov parametrization.}
\label{tab:tabpriors1}
\end{center}
\end{table*}
\twocolumngrid

\section{Numerical analysis}
\label{sec5}

In the following, we will analyze numerically both parametrizations using MCMC  methods. 
\subsection{Mukhanov Parameterization}
The Mukhanov scenario is described by:
\begin{equation}
\label{parameter}
 \{\omega_b, \omega_c, \Theta_s, \tau, \log[10^{10}A_{s}],  \alpha, \beta, N_\star\}~,
\end{equation}

\noindent with $\omega_b\equiv\Omega_bh^{2}$ and $\omega_c\equiv\Omega_ch^{2}$ the physical baryon and cold dark matter energy densities respectively, 
$\Theta_{s}$ is the ratio between the sound horizon and the angular
diameter distance at decoupling, $\tau$ is the reionization optical depth, 
$A_{s}$ the amplitude of the primordial spectrum and $\alpha$ and $\beta$ are the parameters governing the Mukhanov parameterization. For the sake of simplicity, we have assumed that the dark energy component is described by a cosmological constant. Table~\ref{tab:tabpriors1} specifies the priors considered on the cosmological parameters listed above. Notice that this analysis is different from the ones presented in Ref.~\cite{Barranco:2014ira}, as we are also varying here the number of e-folds $N_\star$ to compute the inflaton excursion.  The commonly used $(r, n_s)$ parameters can be easily recovered using Eqs.~(\ref{eq:nsr}),  and the running for this inflationary scheme is completely fixed, see \eg~\cite{Mukhanov:2013tua,Barranco:2014ira}.  The field excursion is computed using Eq.~(\ref{eq:field}).  In our analysis, we also assume the so-called inflation consistency relation ($n_t = -r/8$)  which still holds in the Mukhanov phenomenological model~\footnote{For  recent cosmological analyses relaxing this condition, see Ref.~\cite{Cortes:2014nqa}.}. In order to compute the allowed regions in the derived parameter spaces $(r, n_s)$ and $(r, \Delta \phi)$, we make use of the CAMB Boltzmann code~\cite{camb}, deriving posterior distributions for the cosmological parameters by means of a MCMC analysis, performed using \texttt{CosmoMC}~\cite{Lewis:2002ah}. 

\begin{figure}[!t]
\vspace{-0.55cm}
\hspace*{-0.8cm}
\includegraphics[width=0.57\textwidth, height=8cm]{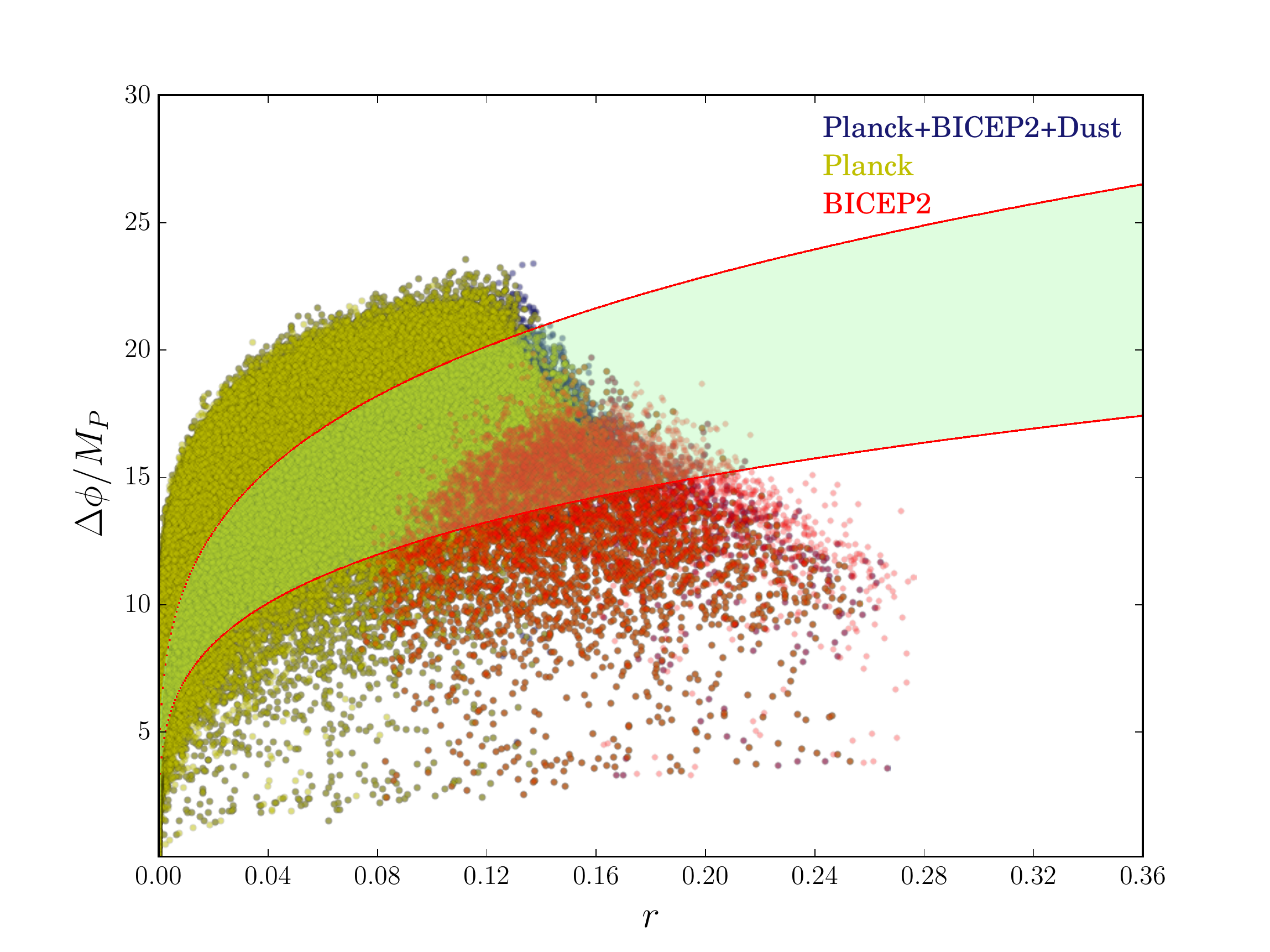}
\caption{The result of the Monte Carlo reconstruction of $6\times 10^6$ inflationary models using {\tt Flowcode1.0}, represented in the $(r, n_s)$ plane. Here, the light green area represents the theoretical prediction \eq{EM} for $N_\ast=40 - 70$.} 
\label{fig:fig4}
\end{figure}

The basic data set used for our numerical analyses includes the Planck CMB temperature anisotropies data \cite{Ade:2013ktc,Planck:2013kta} together with the WMAP 9-year polarization data~\cite{Bennett:2012fp}.  The total likelihood for the former data is obtained by means of  the Planck collaboration publicly available likelihood code, see Ref.~\cite{Planck:2013kta} for details. The Planck temperature power spectra reaches a maximum multipole  number $\ell_{\rm max}=2500$,  while the WMAP 9-year polarization data is analyzed up to a maximum multipole $\ell=23$~\cite{Bennett:2012fp}. We shall refer to the basic data set in the following as CMB data.

We have also considered the BICEP2 measurements of the tensor-to-scalar ratio $r=0.2^{+0.07}_{-0.05}$~\cite{bicep2,bicep22}. These measurements are included in our analysis by post-processing the chains that were previously generated, using the likelihood code released by the BICEP2 experiment, including the $9$ bandpowers from multipoles $\ell \sim 45$ to $\ell \sim 300$. The recent estimates of the galactic dust polarized emission carried out by the Planck collaboration in Ref.~\cite{Adam:2014bub} have also been included in our numerical fits. For the former purpose, we have added the dust power spectrum measured by Planck in the $40< \ell<120$ multipole range,  $D^{BB}_{\ell} \equiv
\ell (\ell+1)/2\pi C^{BB}_\ell= 1.32 \times 10^{-2}\mu$ K$^2$, to the theoretical $B$-mode spectra in the same multipole range, in order to evaluate the likelihood of the total signal resulting from the addition of gravitational lensing, primordial $B$-modes, and dust $B$-mode contributions. The statistical and the interpolation-induced uncertainties of the Planck dust analysis are accounted for by including them in the BICEP2 covariance matrix. We then use this Planck dust plus BICEP2 likelihood to postprocess the chains previously obtained by the Planck temperature and WMAP9 polarization likelihoods. We multiply the original weight of each model by the Planck dust plus BICEP2 likelihood, using the new weights to derive the allowed cosmological parameter regions by Planck CMB data, Planck dust polarization measurements and BICEP2.

In Fig.~\ref{fig:fig1}, we plot the $68\%$ and $95\%$ confidence regions in the plane of the \emph{derived} parameters $n_s$ and $r$. We also  superimpose the region covered by the Mukhanov parametrization for $40\le N_\ast\le 70$, see Eqs.~(\ref{eq:ns})  and (\ref{eq:r}).  We represent the MCMC results for the three possible data combinations. Notice that CMB data alone shows a mild preference for the Branch I region (with a negligible tensor-to-scalar ratio $r$), since there is no $68\%$~CL allowed contour in the Branch II region. The inclusion of BICEP2 measurements to CMB data isolates the Branch II region as the allowed one at $95\%$~CL, favoring inflationary scenarios with a relatively large tensor-to-scalar ratio, like for instance chaotic inflationary models. However, once that the galactic polarized dust emission from the Planck experiment is taken into account in the BICEP2 likelihood, there is no difference between Branch I and Branch II regions, as both regions are equally allowed by the data.   

Figure \ref{fig:fig3} shows the $68\%$ and $95\%$ CL allowed regions in the plane of the \emph{derived} parameters $r$ and $\Delta \phi$. As previously stated, to derive $\Delta \phi$, we have used  Eqs.~(\ref{eq:field}). We also plot the theoretical relationship \eq{EM}, for $40\le N_\star \le 70$. Notice that the area covered by this relationship  perfectly agrees with the parameter regions preferred by current cosmological data.  Notice as well that CMB data alone favours relatively small inflaton excursions, as this is the expected behaviour in scenarios in which $r$ is tiny, like for instance in Starobinsky models, belonging to Branch I. The inclusion of BICEP2 data favours instead large inflation  excursions \ie $\Delta \phi/M_P \sim 20$,  at $95\%$~CL. Such large excursions have been argued to render the validity of effective field theory questionable. In this regime, non-renormalizable operators ${\cal O}_{n+4}=c_n \phi^{n+4}/M_P^n$ are expected to dominate the inflationary potential, compromising its flatness, even in the regime of validity of classical general relativity $V\ll M_P^4$. Suppressing such operators is only possible if the shift symmetry $\phi\to \phi+c$ is {\it only} broken softly at the renormalizable level. However, since in general this symmetry is a mere global symmetry, it is likely to be badly broken by gravity, producing the non-renormalizable operators ${\cal O}_{n+4}$. Furthermore, embedding the theory in a framework where shift symmetry descends from a local symmetry leads to inconsistencies \cite{Banks:2003sx}.  

However, for sub-Planckian inflaton excursions the problems discussed above are less severe. Fortunately, once the Planck dust polarization measurements are included in the analyses together with CMB and BICEP2 data, the small excursion region becomes allowed at $95\%$~CL and therefore trans-Planckian field values are no longer absolutely required to explain observations. This is the second main result of this study.

\subsection{The Hubble Flow Formalism}

We have performed as well an analysis of the $6\times 10^6$ models resulting from integrating the Hubble flow equations, using the priors Eqs.~(\ref{intervals}). For each of these models, we have computed the likelihood by means of the covariance matrices resulting from three different MCMC runs with flat priors in $n_s$, $r$ and $\alpha_s$~\footnote{The authors of Ref.~\cite{Contaldi:2013mua} performed a MCMC analysis considering the Hubble flow parameters as free parameters, deriving constraints on $n_s$, $r$ and $\alpha_s$. However, the resulting  cosmological constraints on these derived parameters are not significantly affected, and their bounds were similar to those found in the case in which the parameters $n_s$, $r$ and $\alpha_s$ are free parameters in the Monte Carlo. Therefore, we shall use the likelihood in terms of  $n_s$, $r$ and $\alpha_s$ rather than in terms of the Hubble flow parameters.}. The former three runs correspond to the three possible data combinations considered in this study, namely, CMB data alone, CMB plus BICEP2 measurements, and finally, CMB plus BICEP2 plus Planck dust polarization measurements.  The covariance matrices were previously marginalized over the remaining cosmological parameters that are irrelevant for our purposes.   

Figure \ref{fig:fig4} shows the analogue of Fig.~\ref{fig:fig3} but for the Hubble flow analysis  in the $(r,\Delta \phi)$ plane. The models depicted are allowed at the $95\%$~CL  by the three different data sets.  We also include in Fig.~\ref{fig:fig4}  the theoretical prediction from Eq.~(\ref{EM}) for $40\le N_\star \le 70$. Notice that the allowed regions for the inflationary Hubble flow approach almost coincide with those arising from the Mukhanov parametrization, and consequently these two approaches are equivalent from the point of view of data analyses.

\section{Discussion and Conclusions}
\label{sec6}
Unraveling the source of primordial curvature perturbations is one of the key purposes of modern cosmology, both from the theoretical and observational viewpoint. The inflationary paradigm is the leading mechanism that provides such initial conditions. In this regard, when testing the inflationary predictions against cosmological measurements, the approach used to describe inflation is crucial. The most familiar picture is based on the dynamics of a friction-dominated scalar field. However, this description, although useful,  is always model-dependent as the  predictions for the cosmological observables will largely depend on the inflationary potential. Furthermore, when embbeded in a  consistent fundamental theory, the shape of this latter is usually difficult to understand. In this work, we focused on two model-independent approaches, that might alleviate the above problems. The first one is a pure theoretical formulation, the Mukhanov parametrization, in which inflation is described via an effective equation of state. The second approach is a pure phenomenological one, which deals with the reconstruction of the inflationary trajectory via the slow-roll hierarchy.  We showed that the allowed parameter regions arising from fitting these two approaches to current CMB data (temperature and polarization) agree with the expected fixed-point solutions. Remarkably, the parameter regions recovered from both model-independent methods are almost identical. Our results thus suggest that these two approaches are the most suitable ones to constrain the inflationary parameters, as they are independent of the inflaton potential details while ensuring a successful inflationary period. 

Another problem that we touched upon in this work is the issue of super-Planckian inflaton field values. Such large excursions have been argued to cause the breakdown of effective theories (see \eg \cite{Conlon:2012tz, Boubekeur:2013kga}). At small inflaton values, the effective theory approach makes sense, and no additional fine-tuning is required to make the potential flat. However, once the inflaton reaches super-Planckian values, it is really difficult to justify the absence, or at most the extreme suppression, of higher order non-renormalizable terms in the inflaton potential, without the knowledge of a UV-complete theory. The BICEP2 collaboration \cite{bicep2} has claimed the detection of $B$-modes on large scales. If the primordial nature of this signal is confirmed, then it would constitute an unmistakable smoking gun of inflation. Furthermore, the amplitude of the detected signal suggests that, if we insist on describing inflation as a scalar field dynamics, then the regime of super-Planckian excursions should be consistently understood. In this work, we have reconstructed the inflaton excursion using the two approaches described above. Our analyses indicate that the inflaton excursions required to explain the BICEP2 data can take sub-Planckian values once the galactic dust polarized signal measured by Planck is accounted for. As a consequence, the validity of effective field theories to describe inflation as a scalar field dynamics still holds. The forthcoming polarization data release from the Planck collaboration will fortunaltely shed light on this crucial issue. 
\section{Acknowledgments}
O.M. is supported by the Consolider Ingenio project CSD2007-00060, by
PROMETEO/2009/116, by the Spanish Ministry Science project FPA2011-29678 and by the ITN Invisibles PITN-GA-2011-289442. We also thank the Spanish MINECO (Centro de excelencia Severo Ochoa Program) under grant SEV-2012-0249.

\end{document}